# Silicene on Substrates：A Theoretical Perspective

Zhong Hongxia (钟红霞),[1, 3†] Quhe Ruge (屈贺如歌),[1,4, 5,6,†] Wang Yangyang (王洋洋),[1,7, †] Shi Junjie (史俊杰),[1] and Lu Jing (吕劲)[1,2,*]

[1]State Key Laboratory for Mesoscopic Physics and Department of Physics, Peking University, Beijing 100871, P. R. China

[2]Collaborative Innovation Center of Quantum Matter, Beijing 100871, P. R. China

[3]Department of Physics, Washington University in St. Louis, St. Louis, Missouri 63130, USA

[4]Academy for Advanced Interdisciplinary Studies, Peking University, Beijing 100871, P. R. China

[5]State Key Laboratory of Information Photonics and Optical Communications, Beijing University of Posts and Telecommunications, Beijing 100876, China

[6]School of Science, Beijing University of Posts and Telecommunications, Beijing 100876, China

[7]Department of Nuclear Science and Engineering and Department of Materials Science and Engineering, Massachusetts Institute of Technology, Cambridge, Massachusetts 02139 , USA

[†]These authors contributed equally to this work.

Email: jinglu@pku.edu.cn

## Abstract

Silicene, as the silicon analog of graphene, has been successfully fabricated by epitaxial growing on various substrates. Similar to free-standing graphene, free-standing silicene possesses a honeycomb structure and Dirac-cone-shaped energy band, resulting in many fascinating properties such as high carrier mobility, quantum spin Hall effect, quantum anomalous Hall effect, and quantum valley Hall effect. The maintenance of the honeycomb crystal structure and the Dirac cone of silicene is crucial for observation of its intrinsic properties. In this review, we systematically discuss the substrate effects on the atomic structure and electronic properties of silicene from a theoretical point of view, especially focusing on the changes of the Dirac cone.

## Introduction

As the silicon analogy of graphene, silicene shares graphene's both outstanding properties like the Dirac-cone-shaped energy band[1] and the high carrier mobility, theoretically of the same order of $10^3$-$10^5$ $cm^2$/ (V s),[2, 3] and its zero band gap nature.[1] Unlike planar graphene, whose band gap is difficult to open without degrading its electronic property,[4] a tunable band gap can be opened in low-buckled silicene (0.44 Å) by a vertical electric field,[5, 6] single-side surface adsorption of alkali and transition metal atoms without degrading the electronic properties of silicene.[7, 8] Meanwhile, three (*p/i/n*) doping types are realized for silicene sheets, which can be used to fabricate silicene-based *p–i–n* transport field effect transistors (FETs).[8] Because of the high carrier mobility, a tunable band gap, and easy integration into the modern Si-base device technologies, silicene is a promising material for high-speed electronic devices, and the silicene-based transistor will own an extremely fast speed, making it able to operate in the THz frequence range.[9] Very recently, a silicene FET operating at room temperature has been realized via growth-transfer-fabrication process, corroborating the expected ambipolar Dirac charge transport.[10]

Due to the stronger spin-orbital coupling (SOC) of Si atom than that of C atom, silicene with a SOC band gap about 1.55 meV (vs $24\times10^{-3}$ meV in graphene)[11] is more eligible in realizing quantum spin Hall effect (QSHE) and fabricating spintronics devices.[12] By adjusting an electric field and/or an exchange field, QSHE, quantum anomalous Hall effects (QAHE) and valley-polarized QAHE, quantum valley Hall effects (QVHE), and valley polarized metal (VPM) state can be realized in silicene,[13, 14] providing many potential applications in spintronics and valleytronics. Based on these topological states, spin-filters and spin/valley separators are designed with quite good performance.[15] Therefore, silicene serves as an ideal platform for efficiently manipulating spin/valley degrees of freedom.

Unfortunately, silicene is much less stable and more difficult to synthesis than graphene. It is well known that graphene, with $sp^2$-bonded carbon atoms, can be exfoliated from graphite easily because of the weak van der Waals (vdW) interlayer interactions of graphite. However, silicene, a low-buckled structure with partial $sp^3$-bonded silicon atoms, cannot be exfoliated from bulk silicon due to the strong covalent Si-Si bonds. Thus, the exfoliation technique developed for graphene does not work for silicene. Hence the major method for synthesizing silicene sheet is epitaxial growth on solid surfaces. Silicene has been first successfully synthesized by silicon deposition onto Ag(111) substrates,[16, 17] and further obtained by segregation on top of $ZrB_2$ thin films epitaxial growth on Si wafers,[18] and then by directly depositing silicon atoms onto Ir(111) surface and annealing the sample at 670 K,[19] and deposited on the ZrC(111) at 800 K.[20] Recently, high-buckled two dimensional (2D) Si nanosheets (2 Å) has been fabricated by molecular beam epitaxial growth on bulk $MoS_2$ surface.[21] Many other substrates, including graphene, Ca, Pb, SiC, BN, $MoX_2$ (X = S, Se, and Te), $WSe_2$, and GaS, have been proposed and investigated theoretically.[22-27]

In order to observe the high carrier mobility and various topological properties, the honeycomb crystal structure and the Dirac cone of silicene have to be maintained. Therefore, the effects of the substrates on the structural and electronic properties of silicene must be clarified. In this review, we will discuss the properties of silicene on various substrates. Special attention will be paid on whether the key Dirac cone of free-standing silicene can be preserved on substrate. The reviewed substrates generally satisfy two criteria: hexagonal symmetry lattice and a small lattice mismatch ($\Delta a$) with silicene. This review is organized into the corresponding two sections as follows. In the first section, we discuss the electronic properties of epitaxial silicene layers on Ag, $ZrB_2$ (ZrC), Ir, and other metallic substrates. Due to enhanced band hybridization, the absence of the Dirac cone is a common character for silicene on metal substrates, except for graphene, Ca and Pb. On these three metal surfaces, supported-silicene can preserve its intrinsic Dirac cone below the Fermi level ($E_f$). In the second section, silicene on insulating and semiconducting 2D layered substrates, including SiC, BN, $MoX_2$ (X = S, Se, and Te), $WSe_2$, GaS, and other non-metallic substrates have been summarized. It is found that the lattice mismatch and work function play crucial role in the electronic properties of silicene on non-metallic substrates.

## 1. Silicene on metallic substrates

Because silicene does not exist in nature, we must evaporate and deposit Si atoms on substrates (or templates) to synthesize it. The major method for synthesizing silicene is epitaxial growth on metallic substrates up to now. For instance, silicene has been successfully grown on metallic Ag, $ZrB_2$,[18] Ir[19], and ZrC[20] substrates. Additionally, silicene has to be contacted with metal electrodes in real device applications. Understanding the nature of silicene on metallic surfaces will pave a way for investigating other important physical phenomena such as superconductivity.[28-30] Many efforts thus have been put to both theoretical and experimental researches of silicene on metals. Based on the periodic table, we focus on the researches of fcc metals with (111) surfaces and hcp metals with (0001) surfaces because of their hexagonal symmetry, which are labeled as the brightness in Fig. 1.[22]

### 1.1 Silicene on Ag(111) surface

Among all the metallic substrates, Ag is the most common substrate to synthesis silicene experimentally because of the moderate and homogeneous interaction, and lattice match between Ag(111) surface and silicene.[31] Several phases of silicene have been successfully fabricated by epitaxial growth on Ag(111) with superstructures of $(3\times3)$silicene/$(4\times4)$Ag(111),[16, 32-35] $(2\times2)$silicene/($\sqrt{7}\times\sqrt{7}$)Ag(111),[32] ($\sqrt{7}\times\sqrt{7}$)silicene/($2\sqrt{3}\times2\sqrt{3}$)Ag(111),[17, 32, 35, 36] ($\sqrt{7}\times\sqrt{7}$)silicene/($\sqrt{13}\times\sqrt{13}$)Ag(111),[32, 34, 35] and $\sqrt{3}\times\sqrt{3}$ silicene,[37, 38] as shown in Fig. 2.[32] The appearance of different superstructures is attributed to the different substrate temperatures and deposition rates.[35] Among these different superstructures, the phase of $(3\times3)$silicene/$(4\times4)$Ag(111) has attracted most attention due to the small $\Delta a$ (0.3%). Experimentally, there is a fierce debate over the existence of Dirac cone in silicene on Ag surface. The scanning tunneling microscope (STM) and angle-resolved photoemission spectroscopy (ARPES) results in phases $(3\times3)$silicene/$(4\times4)$Ag(111) provide an evidence that both structural and electronic properties of the 2D silicene are very similar to those of graphene.[37] Unfortunately, the scanning tunneling spectra (STS) in $(3\times3)$silicene/$(4\times4)$Ag(111) and ($\sqrt{7}\times\sqrt{7}$)silicene/($\sqrt{13}\times\sqrt{13}$)Ag(111) phases demonstrate that the two Ag-supported silicene phases have lost their Dirac fermions and 2D characters, because no Landau level (LL) sequences peculiar to Dirac electrons appear under a strong magnetic field (Fig. 3c) as the LL observed in the highly ordered pyrolytic graphite (Fig. 3d).[33] Subsequently, extensive theoretical calculations clarify that the Dirac cone in the observed phases of $(3\times3)$silicene/$(4\times4)$Ag(111), $(2\times2)$silicene/($\sqrt{7}\times\sqrt{7}$)Ag(111),

($\sqrt{7}\times\sqrt{7}$)silicene/ ($2\sqrt{3}\times2\sqrt{3}$)Ag(111), ($\sqrt{7}\times\sqrt{7}$)silicene/($\sqrt{13}\times\sqrt{13}$)Ag(111) is severely destroyed by the strong hybridization between silicene and Ag substrates in Figs. 4-5.[33, 39-46] The experimental observed Dirac-cone-like features in (3×3)silicene/(4×4)Ag(111) and ($\sqrt{7}\times\sqrt{7}$)silicene/($\sqrt{13}\times\sqrt{13}$)Ag(111) phases are ascribed to the *s-p* bands of bulk Ag or the silicene-Ag hybridization instead of the intrinsic bands of silicene.

The situation becomes more complicated for the newly found phase of ($\sqrt{3}\times\sqrt{3}$) silicene on Ag(111) surface. Chen *et al.* confirms the existence of Dirac cone in the phase of ($\sqrt{3}\times\sqrt{3}$) silicene/Ag(111) from the quasiparticle interference patterns (QIP) obtained by STS experiments.[37] The flip-flop motion between the two degenerate ($\sqrt{3}\times\sqrt{3}$) structures explains their hexagonal symmetry STM images at higher temperature and its freeze at lower temperature.[38] However, the evaluated energy barrier for the flip-flop motion (38 meV per Si atom by VASP),[38] is too large to explain the frozen temperature 40 K. Chen *et al.* also argued that the vdW force is crucial to obtain the rhombic ($\sqrt{3}\times\sqrt{3}$) structure,[38] which is in contrast with the covalent bonds between silicene and Ag in other phases.[42, 43] Moreover, the band structure in Chen's paper ignores the interaction between silicene and Ag.[38] By contrast, Guo *et al.* argue that the phase ($\sqrt{3}\times\sqrt{3}$)silicene/Ag is bilayer silicene instead of monolayer (ML) silicene.[47, 48] It is the second layer that shows the linear energy dispersions, deduced from the atomic scale STM and on-contact atomic force microscopy (nc-AFM) images.[47, 48] This is verified by later theoretical calculations,[49] which indicates that the dispersion (vdW) interactions only play a minor role in the ML silicene/Ag(111) and the energy barriers among the three stable ($\sqrt{3}\times\sqrt{3}$) structures are in the range of 7-9 meV per Si atom. The linear energy dispersion reported in the Chen's experiments[37] corresponds to the silicene-Ag orbital hybridization and shift downwards deep in the valence band.

Although it is controversial whether the Dirac cone is present for Ag-supported silicene in experiments, most theoretical calculations reach the consistent conclusion that the Dirac cones of all Ag-supported silicene are severely destroyed by the strong band hybridization between silicene and Ag substrates from the electronic band structure. In Table 1, we present the theoretical structural parameters of hybrid structures, such as supported-silicene buckling ($\Delta$), the vertical distance from the bottom silicene layer to the topmost metal layer ($d_0$), binding energy ($E_b$) (per Si atom) required to remove the silicene sheet from the surface, and charge transfer ($Q$). The various phases of Ag-supported silicene display large Δ of 0.95~1.65 Å, small $d_0$ ranging from 1.23~1.87 Å, $E_b$ of 0.7 eV, and $Q$ of -0.07|$e$|.[40, 46, 49]

### 1.2 Silicene on $ZrB_2$(0001) and ZrC(111) surface

With high melting point, high hardness, high stability to oxidation, and high electric conductivity, transition-metal carbides (TMCs) and diborides belong to a group of compounds, which have promising applications to high-temperature ceramics and are highly stable substrates for film growth. Fleurence *et al.* successfully synthesized the phase ($\sqrt{3}\times\sqrt{3}$)silicene/(2×2)$ZrB_2$(0001) through surface segregation on zirconium diboride thin films grown on Si wafers via epitaxial growth.[18] In contrast to the planar and buckled forms of free-standing silicene, the phase ($\sqrt{3}\times\sqrt{3}$)silicene/(2×2) $ZrB_2$(0001) is “planar-like” phase, with all but one Si atoms per hexagon residing in the same plane.[50, 51] The plane is characterized by a residual buckling of just 0.01 Å, with the remaining atom protruding at a height of 1.58 Å. This indicates there is a strong interaction between epitaxial silicene and substrate $ZrB_2$, so that the Dirac point epitaxial silicene on $ZrB_2$ is destroyed by the hybridization of Si $sp^2$ and $p_z$ orbitals, accompanied by the Zr $d$ orbitals.[50, 51] Apart from $ZrB_2$, Aizawa *et al.* has fabricated silicene on ZrC(111) in the same phase ($\sqrt{3}\times\sqrt{3}$)silicene/(2×2) ZrC(0001) with more perfect Δ$a$ (1%) compared with the value (5%) for $ZrB_2$ case.[20] Similar to the silicene on $ZrB_2$, the phase ($\sqrt{3}\times\sqrt{3}$) silicene/(2×2) ZrC(111) is also “planar-like” one, with a residual plane buckling of just 0.01 Å but the remaining atom protruding at a height of 1.35 Å . No crossing of the π band is observed on $E_f$, suggesting the absence of expected Dirac cone because of the considerable covalent nature for ($\sqrt{3}\times\sqrt{3}$) silicene/(2×2) ZrC(0001) from theoretical calculations.[20]

### 1.3 Silicene on Ir(111) surface

Silicene has been successfully fabricated on Ir(111) surface by directly depositing silicon on the Ir(111) surface and annealing the sample at 670 K.[19] Both theories and experiments verify the 2D continuity and buckled structure of silicene on Ir(111).[19, 52] The phase ($\sqrt{3}\times\sqrt{3}$) silicene/ ($\sqrt{7}\times\sqrt{7}$)R19.1° Ir(111) possesses Δ of 0.83 Å, larger than that (0.44 Å) of free-standing one, $d_0$ of 2 Å, $E_b$ of 1.60 eV, and formation energy of 1.10 eV, indicating the stability and a strong adsorption of silicene on Ir surfaces.[19, 46] The strong adsorption originates from the strong hybridization between S $3p_z$ and Ir $5d$ states through partial density of states (PDOS) analysis.[52] Furthermore, there is a large charge accumulation at the interface between silicene and substrates. Therefore, silicene on Ir(111) loses its original Dirac cone around $E_f$, leading to a metallic band structure.[46, 52]

### 1.4 Silicene on graphene surface

Due to the hexagonal symmetry and similar honeycomb lattice constants, the heterosystem silicene/graphene is reasonable, with no imaginary phonon frequencies.[27, 53-55] Multiple phases of silicene with different orientations relative to graphene have been calculated with small $\Delta a$ (0.5~2.7%).[27] Among the various phases, we take the phase ($\sqrt{3}\times\sqrt{3}$)silicene/ ($\sqrt{7}\times\sqrt{7}$)graphene in Fig. 6 as example. The corresponding projected band structures clearly show that the Dirac cones of silicene and graphene are well preserved, with nearly unchanged Fermi velocities (in the order of $10^5$ m/s) at their Dirac points.[27] More interestingly, the silicene and graphene are weakly *p*- and *n*-doped, respectively, because their Dirac cone slightly shift 0.1 eV above and below $E_f$, in the phase ($\sqrt{3}\times\sqrt{3}$)silicene/ ($\sqrt{7}\times\sqrt{7}$)graphene. The doped characteristics are essentially caused by a small amount of $Q$ ($5\times10^{-4}|e|$) from silicene to graphene, inducing small band gaps for silicene ($\Gamma$) (26 meV) and graphene ($K$) (2 meV).[53] Moreover, a conversion of doping types of silicene and graphene happens when the $d_0$ is above some certain value (4.2 Å for phase (2×2)silicene/(3×3)graphene), resulting from the competition of the electronic states overlap and tunneling energy barrier. Therefore, the effective and tunable self-doping in the heterosystem silicene/graphene provides a great potential for fabricating new *p*-*n* junctions.

Similar to the phase ($\sqrt{3}\times\sqrt{3}$)silicene/($\sqrt{7}\times\sqrt{7}$)graphene, both silicene in-between two graphene layers and silicenen on ML graphene in different supercells possess identical atomic and electronic structures with those in free-standing buckled silicene, as a result of the weak vdW interlay interaction.[54] Meanwhile, the substrate silicene preserves its intrinsic electronic properties well. Therefore, silicene and graphene layers are ideal templates for each other for growth and applications, with a small lattice mismatch and well preserved intrinsic properties.

### 1.5 Silicene on other metallic surfaces

The cases of silicene on other promising metals have been predicated theoretically, including ($\sqrt{3}\times\sqrt{3}$)silicene/(2×2)Mg(0001), ($\sqrt{3}\times\sqrt{3}$) silicene/($\sqrt{7}\times\sqrt{7}$)Al(111), (1×1)silicene/Ca(111), ($\sqrt{7}\times\sqrt{7}$)silicene/($\sqrt{13}\times\sqrt{13}$)Ti(0001), ($\sqrt{3}\times\sqrt{3}$) silicene/($\sqrt{7}\times\sqrt{7}$)Cu(111), ($\sqrt{7}\times\sqrt{13}$)silicene/($\sqrt{13}\times\sqrt{28}$)Zn(0001), (3×3) silicene/($\sqrt{13}\times\sqrt{13}$)Zr(0001), (2×2)silicene/(3×3)Ru(0001), (4×4) silicene/(6×6)Rh(111), (3×3)silicene/($\sqrt{13}\times\sqrt{13}$)Cd(0001), ($\sqrt{3}\times\sqrt{3}$)silicene/($\sqrt{7}\times\sqrt{7}$)Pt(111), ($\sqrt{3}\times\sqrt{3}$) silicene/($\sqrt{7}\times\sqrt{7}$)Au(111), and silicene on Pb(111), as shown in the periodic table (Fig. 1).[22] The absence of Dirac cone due to the strong silicene-Ag hybridization appears in almost all metallic substrates except for Ca and Pb metals.[22, 23, 46]

Let us first discuss the case of silicene on Ca(111) with $\Delta a$ smaller than 1.5%.[22] Among them, the calculated STM images clearly demonstrates the perfect honeycomb symmetry of silicene on 1×1 Ca(111), and a functionalized Ca(1.0 ML)/(1×1) Si(111), accompanied by $\Delta$ of 0.72 Å and 0.69 Å, respectively. Moreover, the Si $p_z$-state derived Dirac cones could be identified near the $K$ point for these two substrates as shown in Fig. 7, with corresponding Fermi velocity in the order of $10^5$ m/s. Furthermore, band gaps (574 and 161 meV) at the Dirac point appear, and the Dirac point is shifted more than 0.5 eV below $E_f$.

The other promising metallic substrate to grow and host Dirac Fermions is Pb(111) surface, due to its chemical stability to Si and its threefold symmetry.[23] Three silicene superstructures are studied with $\Delta a$ no greater than 3.0%. The honeycomb structure of silicene is maintained in all cases, with equivalent $\Delta$ of 0.65~0.74 Å, higher $d_0$ (2.8 Å), lower $E_b$ (0.15 eV), and $Q$ (0.05$|e|$) compared with those for the case of silicene/Ag(111). Consequently, the band structures of the silicene superstructures in Fig. 8 clearly show the original cone-shaped band structures of free-standing silicene, which is located at 0.77 eV below $E_f$, with small gaps of 40~60 meV. The detailed PDOS analysis confirms that the Dirac cone has mainly Si 3$p$ orbitals with very little contribution of Pb 6$p$ states, in contrast with the strong hybridization between Si 3$p$ and Ir 5$d$ states. Therefore, the silicene on Pb is predicted to be a massive or gapped Dirac fermions system, like silicene on Ca.

Except for Ca and Pb metals, the severe destruction of the silicene original Dirac cone due to the strong band hybridization between silicene and metallic substrates appears to be a common character for metal supported-silicene, as shown in Fig. 4 and Figs. 9a-f.[40, 46] Fortunately, the destroyed Dirac cone can be restored by intercalation of low concentration alkali metal atoms between silicene and substrates.[46] After intercalation of K atoms for silicene on Pt, Au, and Al surfaces, the Si atoms are almost in the same plane (Δ= 0 Å), and the zero Dirac cone is well recovered, with the estimated Fermi velocity in the order of $10^5$ m/s, because of the recovered inversion symmetry of the no buckling silicene (Figs. 9g-i). While for the cases of silicene, with intercalating of K atoms, on Ag, Mg, Cu, and Ir surfaces, the recovered Dirac cone is located at 0.40 ~ 0.78 eV below $E_f$, with an opened band gap of 0.15 ~ 0.40 eV, due to the built-in electric filed induced by charge transfer (Figs. 9j-m).

## 2. Silicene on non-metallic substrates

Devices of silicene require non-metallic substrates. Therefore, identifying non-metallic substrates suitable to host silicene is an important issue for the realization of devices based on silicene. We focus on the layered material with a honeycomb structure, such as insulating SiC and BN, transition-metal dichalcogenides $MoX_2$ and GaX (X=S, Se, Te), which are held together by weak vdW. The choice of these hexagonal layered templates for silicene is motivated by the following factors: (1) considerable band gaps (1-6 eV); (2) small lattice mismatch between silicene layer and substrates; (3) pristine surfaces without out-of-plane dangling bonds, may predicting a weak interaction between silicene sheet and substrates.

### 2.1 Silicene on insulating BN/SiC surface

Recent first principles calculations proposed that the outstanding properties like the Dirac cone of silicene can be retained on hexagonal ML boron nitride (*h*-BN) and hydrogenated SiC(0001) surface.[6, 24] The phases of (2×2)silicene/(3×3)BN and (4×4)silicene/(5×5)SiC(0001) are constructed because of their small $\Delta a$ values of 2.04 and 0.08%, respectively. The honeycomb structure of silicene is maintained with a large $d_0$ (about 3 Å) on insulating BN and SiC substrates as shown in Fig. 10. This $d_0$ is comparable to that for graphene/BN system (3.22 Å) and is larger than that for silicene/metals. This large $d_0$ leads to a small $E_b$ of 0.07~0.09 eV, indicating that the interaction between silicene and substrates is rather weak and can be viewed as vdW type. The weak vdW interaction results in small Δ of about 0.50 Å for silicene on BN and SiC, which is very close to the free-standing silicene buckling (0.44 Å). Therefore, the Dirac cone of silicene can be highly expected on these insulating hexagonal templates.

Furthermore, the original linear dispersion band of silicene was clearly preserved on the ML *h*-BN and Si-terminated SiC(0001) surface as shown in Figs. 10e and f. It has to be noted that the gapped Dirac cone is nearly at $E_f$ (without any doping) in the band gaps of the two insulating substrates.[24, 56] By contrast, for the silicene/C−SiC system, the supported-silicene becomes metallic, with the lower segment of the Dirac cone mixing up with the valence bands of the substrate in Fig. 10g. The difference in the band structures of these combined systems is also verified by the partial charge densities analysis, which concentrate on silicene near $E_f$ for silicene/BN and silicene/Si-SiC, but distribute widely into the substrates for silicene/C-SiC.

The physical mechanisms under Fig. 10 have been further explored. The different substrate effects are attributed to the different work functions,[24] *i.e.*, the absolute positions of the valence band maximum (VBM) and conduction band minimum (CBM), of the substrates. The CBM (or VBM) of silicene is slightly 0.1 eV lower than the VBM of C−SiC substrate, so that the charge will transfer from C-SiC substrate to silicene. $E_f$ of the hybrid system is thus higher than that of free-standing silicene, and the silicene in the hybrid system becomes metallic. On the contrary, for the silicene/BN and silicene/Si-SiC systems, the CBM of silicene is located in the gap region of BN and Si-SiC, resulting in little charge transfer between silicene and the substrates. Consequently, the conical point of the Dirac cone in silicene is preserved at $E_f$, with small opened band gaps of 4 meV on BN and 3 meV on Si-SiC.

## 2.2 Silicene on semiconducting $MoX_2$ (X=S, Se, Te) and $WSe_2$ surfaces

Three different possible atomic arrangements of silicene layer (AAA stacking, intermediate position, and ABA stacking) on top of $MoX_2$ (X= S, Se, Te) have been studied.[25, 57, 58] A large $d_0$ (3.0~3.5 Å) and a small $E_b$ (10-100 meV) suggest a vdW interaction between silicene and $MoX_2$. As given in Table 2, as $\Delta a$ decreases from 18% to 9%, $\Delta$ decreases from 1.9 Å for silicene/$MoS_2$ to 0.7 Å for silicene/$MoTe_2$. Thus, the larger the lattice mismatch is, the larger the supported-silicene buckling is. Usually, a large buckling suggests a strong interaction between silicene and substrates, enhancing the possibility of losing Dirac cone. A large $\Delta a$ is also unfavorable for silicene epitaxial growth in experiment.

It is noted that $\Delta$ of silicene on bulk $MoS_2$ is very close to the value for highly buckled free-standing (2.00 Å) silicene structure. Highly buckled silicene is predicted to be metallic with several bands across $E_f$.[1] The energy band structure and the local density of states (Figs. 11a-b) confirm the metallic nature of silicene on bulk $MoS_2$. The gap of $MoS_2$ substrate is preserved, and all the electronic states near $E_f$ are contributed by Si atoms. This clearly verifies that vdW type of silicene/$MoS_2$ interaction, almost without hybridization between silicene and substrates. It is noted that the $\Delta a$ between 4×4 silicene and 5×5 ML $MoS_2$ is rather small (2.42%).[59] In this case, the $\Delta$ of silicene on ML $MoS_2$ is 0.56 Å, close to the value for low buckled free-standing silicene structure. The supported-silicene is slightly *p* doped and basically maintains its linear energy dispersion, as shown in the corresponding electronic band structures and PDOS (Figs. 11c-d). The sublattice symmetry broken induces a small band gap (70 meV) at the Dirac cone. The band gap could be further tuned by an external electric field.

The band analyses are performed for silicene on $MoSe_2$ and $MoTe_2$ in Fig. 11. Interestingly, silicene on bulk $MoTe_2$ in Fig. 11e is a zero-gap semiconductor, preserving its Dirac cones at the *K* point. However, silicene on ML $MoTe_2$ is slight doped by electron and becomes metallic (Fig. 11f), with the band gap of the Dirac cone slightly opened because the inversion symmetry of silicene is destroyed by the charge transfer between silicene and ML $MoTe_2$.[8, 46] Besides, other energy bands also share the same energy with the Dirac cone. In Fig. 11g, silicene is moderately doped by electron on bulk $MoSe_2$, with a larger band gap near the Dirac cone. The geometrical structures and electronic structures for silicene on two $MoX_2$ layers in the same atomic configuration are nearly the same as the cases of silicene on ML $MX_2$.[25]

$WSe_2$ has minimal $\Delta a$ with silicene (0.6%) among the transition metal dichalcogenides.[26] Because of their instability, the combined system varies from metallic to semiconducting with band gap up to 0.3 eV, but accompanied with a well preserved Dirac cone. When $WSe_2$ is doped by S and Te, the combined heterosheet becomes stable, with atomic and electronic structures nearly the same as those for silicene on pristine $WSe_2$. However, high-buckled silicene on $WSe_2$ has a Δ of 1.3 Å, a small $d_0$ of 2.55 Å, and a large $E_b$ of 195 meV, indicating stronger interaction between silicene and substrates compared with the case of low-buckled silicene (Δ of 0.51 Å, $d_0$ of 3.20 Å, and $E_b$ of 120 meV).

### 2.3 Silicene on GaX (X=S, Se, Te) surfaces

GaX (X=S, Se, Te) is another class of chalcogenide compounds with a vdW layered hexagonal structure. Compared with $MoX_2$, the cell parameters are larger for GaX, and the $\Delta a$ (3~8%) with free-standing silicene are smaller.[25, 60] A small $E_b$ (0.07 eV) suggests a weak vdW type of silicene/GaX. The Δ values are small, ranging from 0.30~0.70 Å. The $d_0$ values for silicene on bulk GaX (4.5~5 Å) are much larger than those for silicene on ML GaX (3.1~3.8 Å), similar to the case of $MoX_2$. It appears that the interaction between the silicene and GaX and $MoX_2$ can be enhanced by reducing the layer number.

Fig. 12 displays the electronic band structure of silicene/bulk GaX.[25] Remarkably, both silicene/bulk GaS and silicene/bulk GaSe preserve the linear band dispersion crossing $E_f$ at the $K$ point, with band gaps (0.7 and 1.5 eV, respectively) at the Γ point. These two combined systems are expected to be gapless semiconductors, just like free-standing silicene. Therefore, these heterosheets belong to the Dirac system, possessing outstanding properties like the high Fermi velocity and carrier mobility. Although the Dirac cone is preserved for silicene/bulk GaTe, there is a band touching $E_f$ at the Γ point, and this will mask the intrinsic properties of the Dirac carrier. On condition that the sandwiched configurations have the same stacking modes as the bulk structure, there is nearly no difference between ML GaTe/silicene/ML GaTe (silicene in-between two GaTe layers) and silicene/bulk GaTe. The gapless Dirac cone in silicene/bulk GaS turns into a gapped (about 0.1 eV) one in ML GaS/silicene/ML GaS, due to the interlayer charge distribution. Furthermore, this gap value of ML GaS/silicene/ML GaS can be modulated by an external electric field and strain.[60]

### 2.4 Silicene on other non-metallic substrates ($MgX_2$/AlN/ZnS/$CaF_2$)

Arranged on hexagonal lattice, ML $MgX_2$ (X=Cl, Br, and I) are insulators with calculated band gaps ranging from 3.6~6.0 eV.[61] The lattice mismatches between ML $MgX_2$ and silicene are smaller than 2.9%. Three high symmetry configurations of silicene on ML $MgX_2$(0001) have been checked and turned out to be almost degenerate. All silicene/ML $MgX_2$ heterosheets show small $\Delta$ (0.39 ~ 0.55 Å), large $d_0$ (3.23 ~ 3.75 Å), and small $E_b$ (33 ~ 88 meV), typical of a weak vdW interaction. The band structures of silicene on ML $MgX_2$(0001) for three configurations are displayed in Fig. 13.[61] In each case, the supported-silicene is minor *p* doping and preserves its intrinsic electronic properties well, such as the Dirac-cone-like energy band and high Fermi velocity (estimated $10^5$ m/s). Due to the charge redistribution at the interface of heterosheets, small band gaps are opened in the supported-silicene (1 ~ 90 meV). The opened band gap can be further modulated by an external electric filed and strain, increasing (decreasing) almost linearly with the increasing electric field (strain).[61]

It is known that bulk AlN is wurtzite phase with an experimental semiconducting band gap of 6.5 eV, while ultrathin AlN prefers to hexagonal honeycomb lattice with a smaller energy gap of 4.5 eV. Houssa *et al.* predicted that silicene inserted into ultrathin AlN(0001) stacks preserves its graphene-like electronic properties because of the weak vdW layer interaction and the low silicene buckling (0.22 Å ).[62] Similar to AlN, semiconducting ZnS also is wurtzite phase in bulk and shows hexagonal honeycomb lattice in layers. The forming bonds cause a charge transfer between silicene and ZnS, leading to a semiconducting combined system, with a large opened indirect band gap of 0.7 eV. The Dirac-cone-like energy dispersion is thus unfortunately destroyed.[63] While on $CaF_2$(111), the supported-silicene preserves the original undoped Dirac cone well, with a small opened band gap of 52 meV.[64]

## Conclusion and outlook

In section 1, the electronic properties of epitaxial silicene layers on metallic substrates have been summarized in Table 1. The lattice mismatch ranges from 1~15%, which does not change the conclusion of the existence (absence) of the Dirac cone in metal-supported silicene.

Due to the strong interaction between silicene and most metallic substrates, metal-supported silicene loses its original linear energy dispersion, except for graphene, Ca and Pb substrates. On these three substrates, the weak vdW interaction between silicene and substrates can be obtained, leading to gapped silicene-derived Dirac cone with slightly *n*-doped. For the serious destruction of the Dirac cone in metal-supported silicene, the Dirac cone can be fortunately restored by intercalating low concentration alkali metal atoms between silicene and metallic substrates.

In section 2, we discuss the electronic properties of epitaxial silicene layers on non-metallic substrates as shown in Table 2. Because of pristine substrate surface (no dangling bond), weak vdW interaction between silicene and layered substrates, and appropriate work function, most supported-silicene can preserve its original Dirac cone, with opened small band gap (0~320 meV) and slightly doped. The opened band gap can be further modulated by an external electric field and strain, with the linear band dispersion and high Fermi velocity ($10^5$ m/s). The lattice mismatch and absolute band edge of substrates play vital role in the changing of Dirac cone. Therefore, the hexagonal layered 2D semiconductors and insulators have the advantage to host the Dirac particles compared with the metals.

However, so far low-buckled silicene has been successfully synthesized only on metallic Ag(111), $ZrB_2$(0001), Ir(111), and ZrC(111) templates, which completely destroy the original Dirac cone of silicene due to the strong interaction between silicene and templates in terms of most theoretical and experimental investigations. Weak vdW interaction is thus desired between silicene and templates. Metallic graphene, Ca, Pb and most non-metallic hexagonal layered templates are predicted to host the linear band dispersion of silicene. The lattice mismatch and absolute band edge of substrates play vital roles in the changing of the Dirac cone. Moreover, high-buckled silicene has been successfully synthesized on $MoS_2$ surface very recently,[21] suggesting that non-metallic templates have the potential to grow silicene experimentally. This experimental breakthrough is expected to pave a way to synthesis silicene on 2D layered non-metallic substrates and preserve the intrinsic properties of silicene.

## ACKNOWLEDGMENT

This work was supported by the National Natural Science Foundation of China (No. 11274016/11474012), the National Basic Research Program of China (No. 2013CB932604/ 2012CB619304), and the National Science Foundation Grant (1207141).

Table 1. Structure parameters of the silicene on various metallic substrates. *a* is the lattice constants of materials, and $\Delta a$ is the lattice mismatch between substrates and free-standing silicene. Δ is the calculated buckling of metal supported-silicene. $d_0$ is the equilibrium separation of silicene from various surfaces. $E_b$ is the binding energy (per Si atom) required to remove the silicene sheet from the surface. *W* is the work function for the clean metal surface. *Q* is the Mulliken charge per Si atom transferred from silicene to the metal surfaces. In the last column, Yes (No) stands for the existence (absence) of the Dirac cone of silicene. The lattice constant，low (high) buckling, and work function of free-standing silicene are 3.87 Å, 0.44 (2.00) Å, and 4.48 eV, respectively.

| | *a* (Å) | $\Delta a$ (%) | Δ (Å) | $d_0$ (Å) | $E_b$ (eV) | *W* (eV) | *Q* (eV) | Dirac cone |
|---|---|---|---|---|---|---|---|---|
| Ag[40] | 2.89 | -1.2 | 1.16~1.65 | 1.23~1.87 | 0.41~0.72 | 4.46 | -0.07 | No |
| $ZrB_2$[51] | 3.17 | 5.0 | 1.58 | - | - | 4.61 | - | No |
| ZrC[20] | | 1.0 | 1.35 | - | - | | - | No |
| Ir[46] | 2.71 | 7.2 | 0.83 | 1.95 | 1.69 | 5.47 | 0.13 | No |
| Graphene[27] | 2.45 | -2.0 | 0.62 | 3.30 | 0.07 | 4.50 | - | Yes |
| Ca[22] | 3.92 | 1.5 | 0.69~0.87 | 2.00 | - | 2.87 (bulk) | - | Yes |
| Pb[23] | 3.49 | 1.0 | 0.65~0.74 | 2.80 | 0.17 | 4.25 (bulk) | - | Yes |
| Mg[46] | 3.21 | 4.1 | 1.52 | 1.52 | 0.39 | 3.55 | -0.21 | No |
| Al[46] | 2.86 | 13.2 | 0.29 | 2.12 | 0.35 | 4.06 | -0.06 | No |
| Cu[46] | 2.56 | 1.0 | 1.25 | 1.57 | 0.86 | 4.69 | -0.04 | No |
| Pt[46] | 2.77 | 9.6 | 0.33 | 1.97 | 1.74 | 5.82 | 0.06 | No |
| Au[46] | 2.88 | 14.0 | 0.21 | 1.81 | 0.63 | 5.09 | 0.04 | No |

Table 2. Structure parameters of the silicene on different non-metallic substrates. $E_g$ is the theoretical band gap of the studied substrates (bulk phase), and $W$ is the work function of substrates in ML phase. $E_D$ is the band gap near the Dirac cone. $\Delta E_f$ is the energy shift of Dirac cone with respect to $E_f$.

| | a (Å) | Δa (%) | $E_g$ (eV) | Δ ( Å) | $d_0$ (Å) | $E_b$ (eV) | $W$ (eV) | Dirac cone | $E_D$ (meV) | $\Delta E_f$ (meV) |
|---|---|---|---|---|---|---|---|---|---|---|
| BN[24] | 2.51 | 2.0 | 4.44 | 0.43~0.51 | 3.23 | 0.09 | 5.86 | Yes | 4 | 0 |
| Si-SiC[24] | 3.08 | 0.1 | 2.42 | 0.50 | 2.73 | 0.07 | 5.54 | Yes | 3 | 0 |
| C-SiC[24] | 3.08 | 0.1 | 2.51 | 0.50 | 2.71 | 0.08 | 4.49 | No | - | - |
| $MoS_2$[25, 59] | 3.16 | 18.3(bulk)<br>2.42(ML) | 1.20(bulk)<br>1.84(ML) | 1.90<br>0.56 | 3.00<br>2.93 | 0.10<br>0.12 | 5.12<br>5.36 | No<br>Yes | -<br>70 | -<br>35 |
| $MoSe_2$[25] | 3.30 | 14.7 | 1.10 | 1.00 | 3.0~3.5 | 0.01 | 4.52 | No | - | - |
| $MoTe_2$[25] | 3.52 | 9.0 | 1.00 | 0.70 | 3.0~3.5 | 0.01 | 4.20 | Yes | 0 | 0 |
| $WSe_2$[26] | 3.31 | 0.6 | | 0.51 | 3.20 | 0.12 | 4.37 | Yes | 320 | 150 |
| GaS[25, 60] | 3.58 | 7.5 | 2.50 | 0.70 | 4.5~5.0 (bulk)<br>3.1~3.8(ML) | 0.07 | - | Yes | 170 | 0 |
| GaSe[25] | 3.74 | 3.4 | 2.00 | 0.55 | - | 0.07 | - | Yes | 0 | 0 |
| GaTe[25] | 4.10 | -5.9 | 1.80 | 0.30 | - | 0.07 | - | Yes | 0 | 0 |
| $MgCl_2$[61] | 3.62 | -1.9 | 6.00 | 0.55 | 3.41 | 0.05 | - | Yes | 9~41 | 2~19 |
| $MgBr_2$[61] | 3.83 | 1.0 | 4.80 | 0.49 | 3.23 | 0.06 | - | Yes | 13~64 | 2~24 |
| $MgI_2$[61] | 4.15 | 2.9 | 3.60 | 0.40 | 3.19 | 0.06 | - | Yes | 1~90 | 0~39 |
| AlN[62] | 3.11 | 19.0 | 4.80 | 0.22 | - | - | - | Yes | 0 | 0 |
| ZnS[63] | 3.81 | 2.3 | 3.06 | - | 3.33 | - | - | No | - | - |
| $CaF_2$[64] | 3.88 | 0.5 | - | 0.43 | 2.70 | - | - | Yes | 52 | 0 |

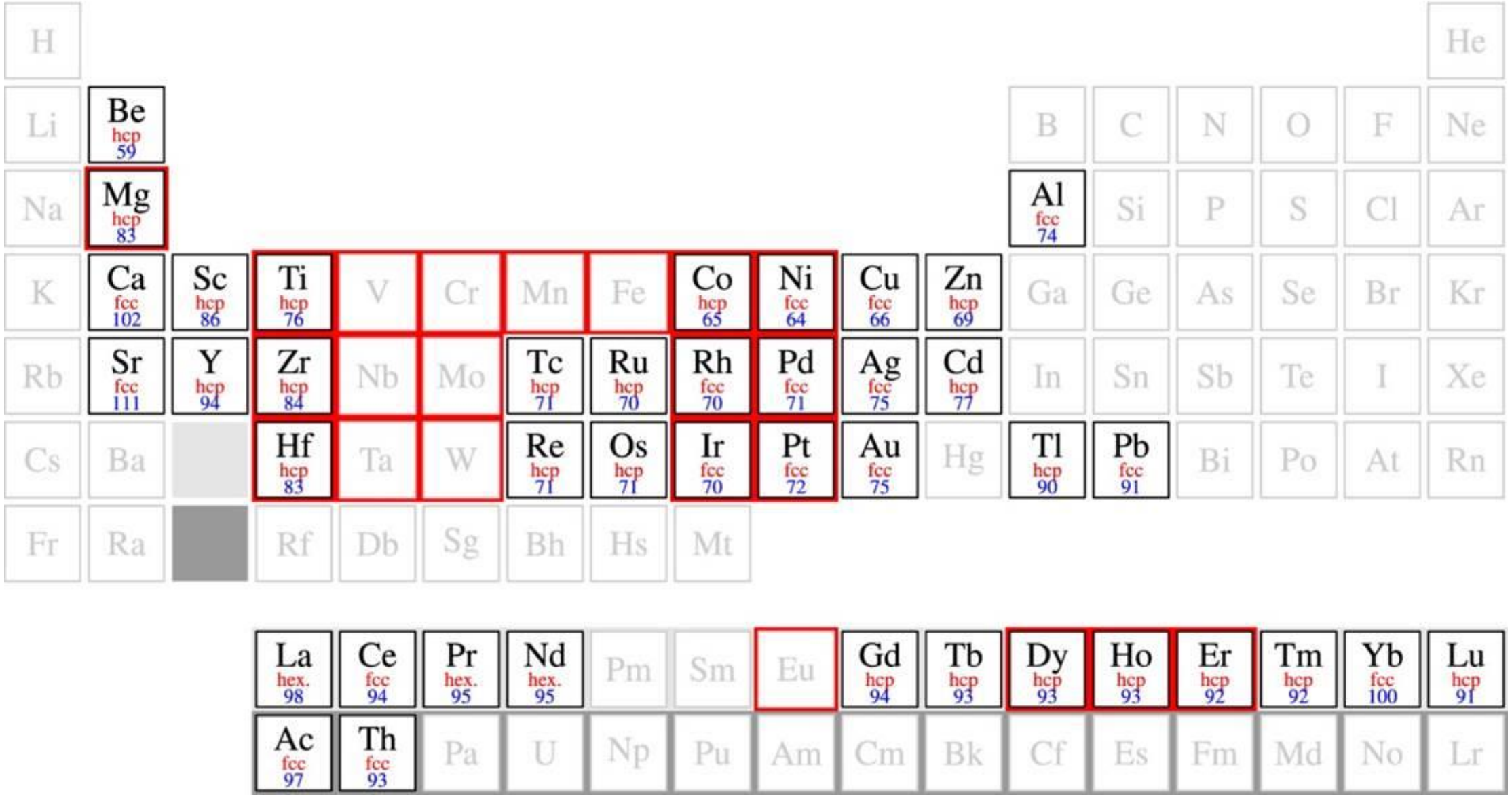

**Figure 1. Periodic table.** The metals with fcc and hexagonal structure are indicated. The corresponding (111) or (0001) surfaces have hexagonal symmetry. The crystal structure is indicated (in red), while the relative lateral lattice constant compared to that of silicene is given as a (blue) number in per cent. The elements in red squares form silicides. Reproduced with permission from Ref.[22]. Copyright 2014 Institute of Physics.

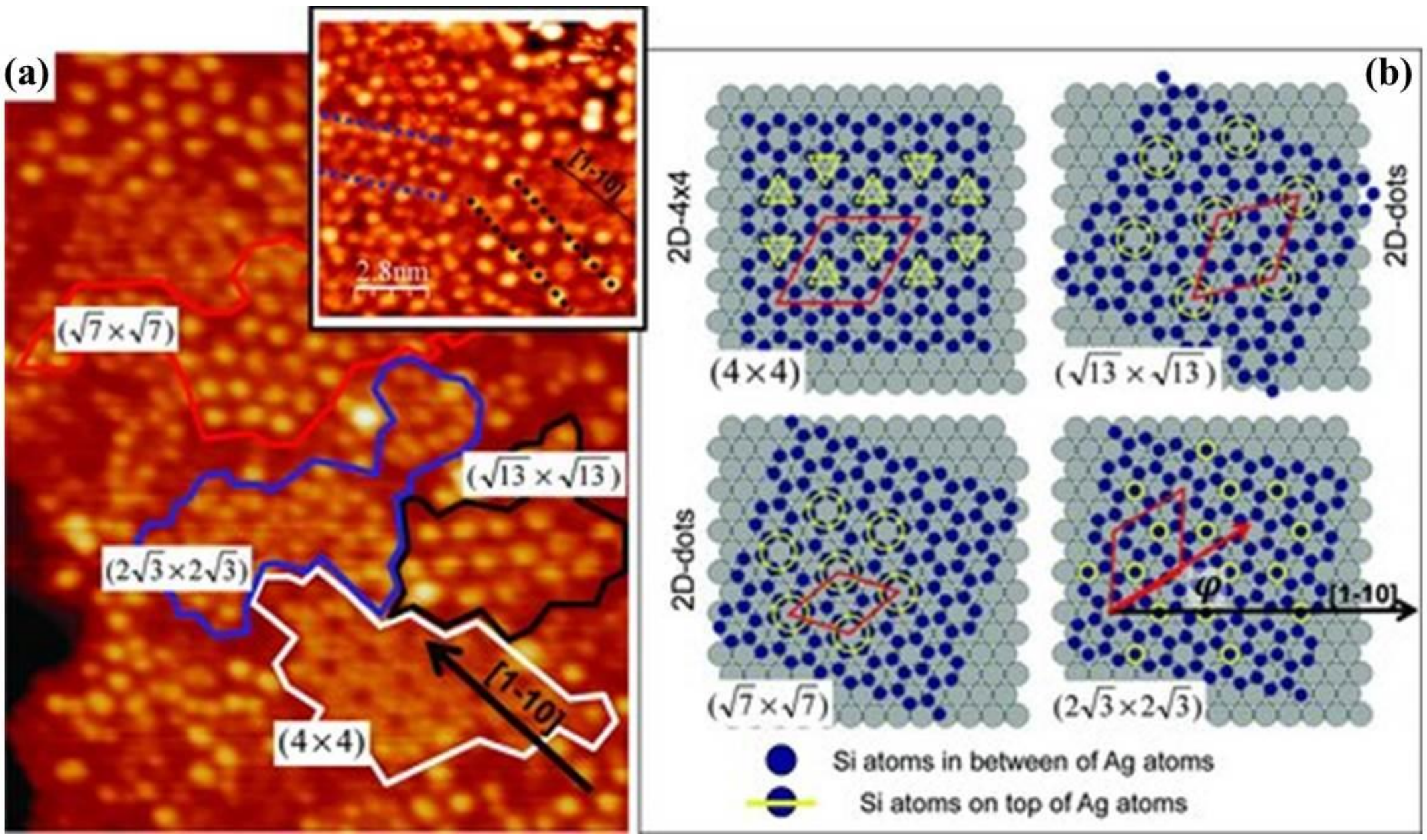

**Figure 2. STM characterization of multi-oriented silicene domains on Ag(111).** (a) From an STM image, a pictorial view of individual Si domains which patch together to form a larger interconnected sheet. In the inset a different STM image (V = –1.5 *V*, I = 0.2 nA) emphasizes the different orientations of the patterned silicene domains; the black arrows indicates the [1–10] direction of the Ag(111) surface, while dotted guidelines (red, blue and black) distinguish the different preferential orientations of the silicene superstructures. (b) Sequence of ball models which simulate the superposition of oriented honeycomb silicene lattices (blue spots) on the Ag(111) surface (grey spots); several superstructures are generated by particular orientations of the silicene with respect to the Ag(111) lattice. Reproduced with permission from Ref.[32]. Copyright 2012 Wiley Online Library.

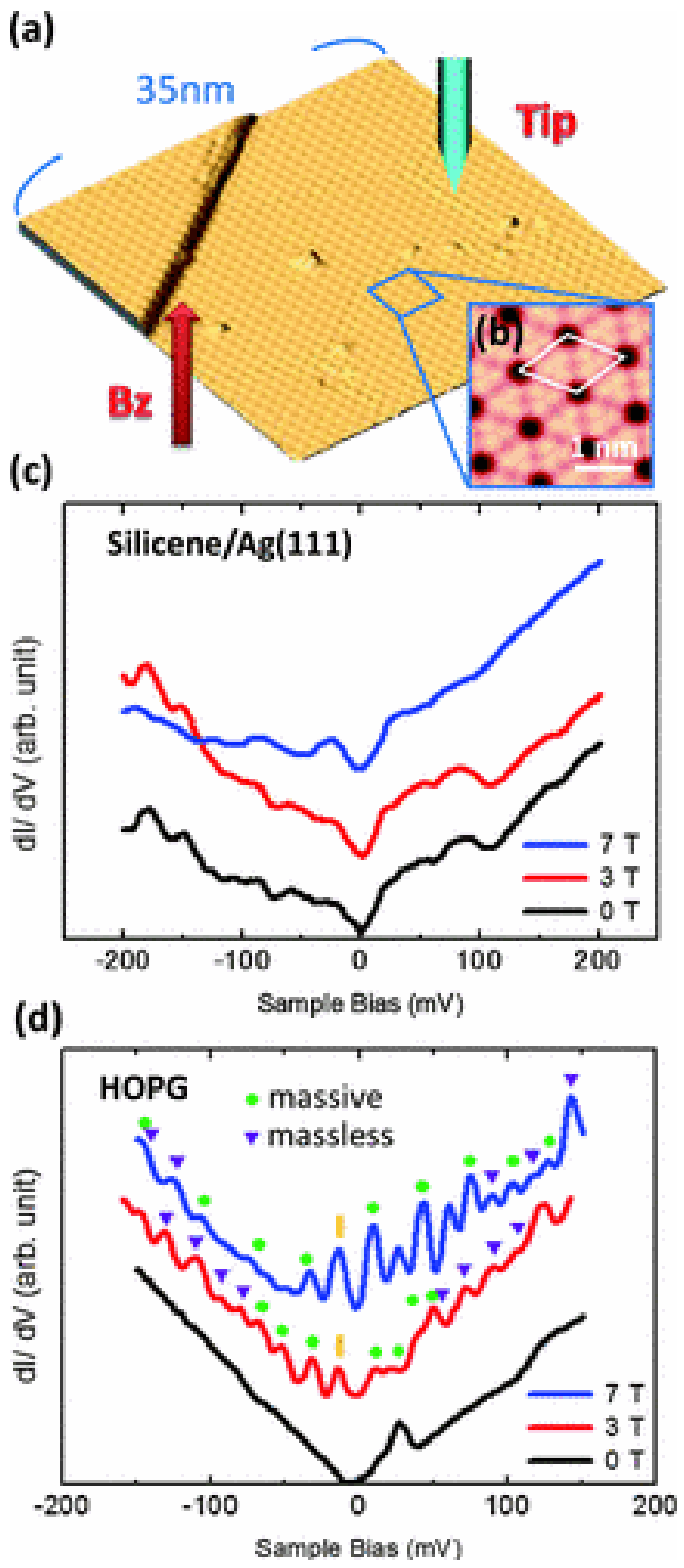

**Figure 3.** (a) STM image of large area of silicene (sample bias voltage $V_s$ = - 0.70 V and tunneling current $I_t$ = 0.19 nA. The image size is 35 × 35 nm$^2$.). (b) High resolution STM image of the 4 × 4 silicene ($V_s$ = + 0.50 V and $I_t$ = 0.30 nA, 3.65 × 3.65 nm$^2$). The unit cell is shown by the white rhombus. (c) The STS spectra of silicene for various magnetic fields perpendicular to the sample surface, $B_z$. (d) The STS spectra of HOPG for various $B_z$. The purple triangles and green circles show the peaks originating from the LLs of massless and massive Dirac fermions, respectively. The $n$ = 0 LL is marked by the yellow bar and the $n$ = 1 LL of massive Dirac fermions is not clearly resolved in 3 T due to low magnetic field. Reproduced with permission from Ref.[33]. Copyright 2013 American Physical Society.

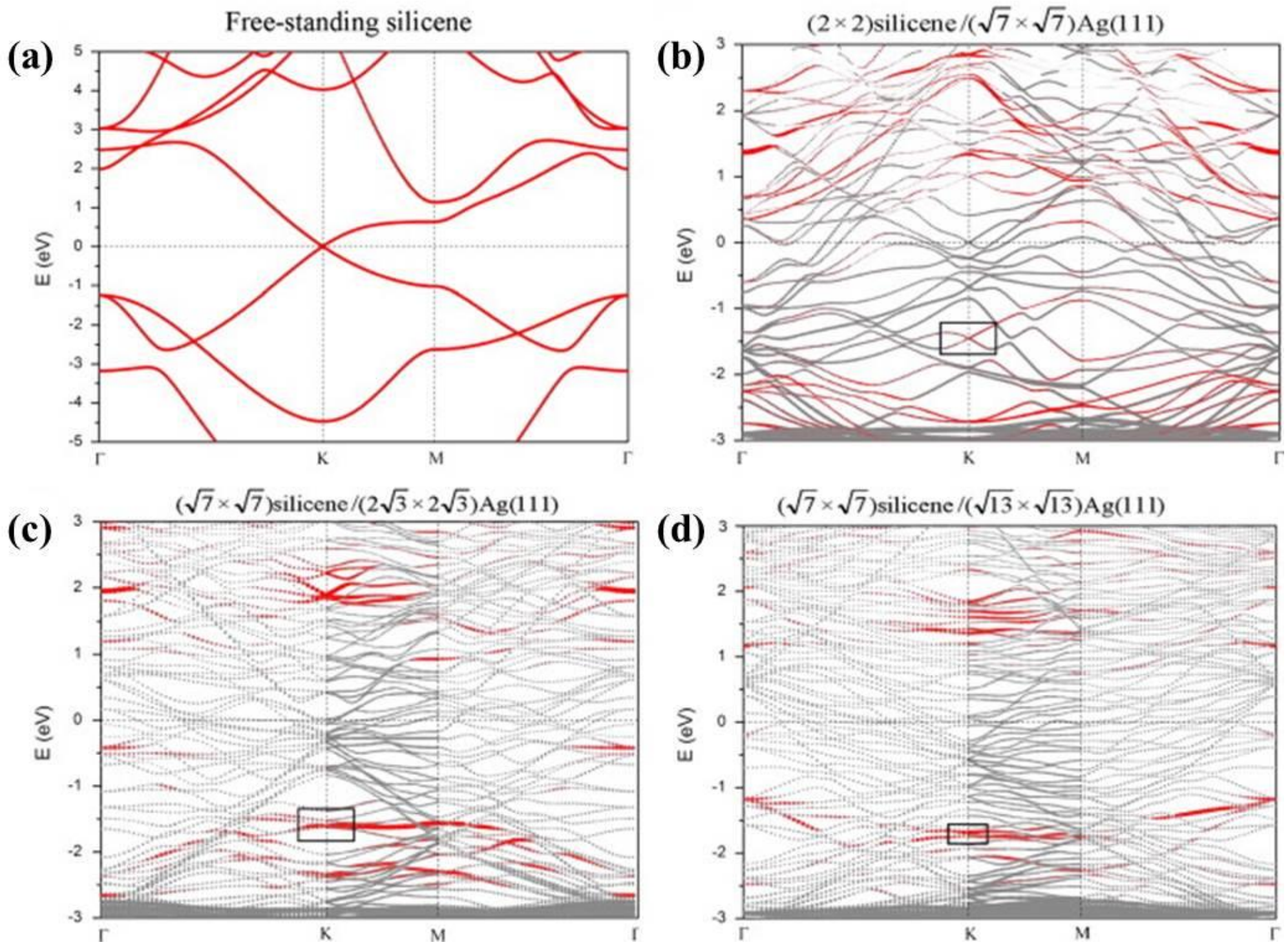

**Figure 4.** (a) Free-standing silicene band structure, in which the π and π* bands degenerate at $E_f$. ((b) and (c)) Projected band structures of silicene/Ag(111) composite systems. Red dots and gray dots represent the states contributed by Ag and Si atoms, respectively.The intensity of the color is proportional to the weight of the corresponding atoms. Black squares indicate those states probably contributed by the π (π*) bands of silicene. $E_f$ is set to zero. (For interpretation of the references to color in this figure legend, the reader is referred to the web version of this article.) Reproduced with permission from Ref.[40]. Copyright 2014 North-Holland.

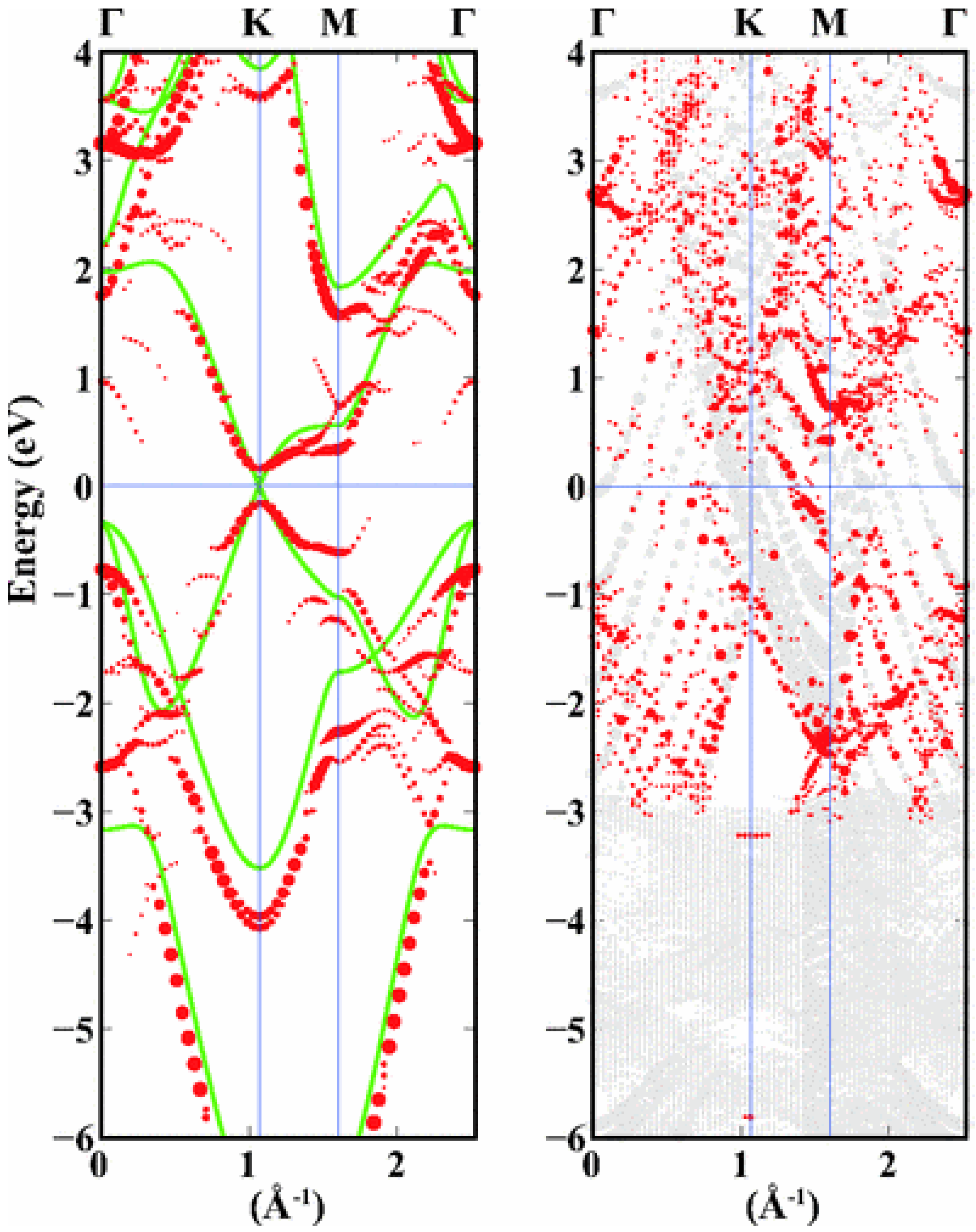

**Figure 5.** Left panel: bands of reconstructed (3×3) silicene in the absence of Ag substrate (unsupported silicene) unfolded to BZ of (1×1) silicene are shown by red dots. The radii of dots correspond to the weight of unfolding. The band structure of ideally buckled silicene is shown by green lines. Right panel: unfolded band structure of silicene on Ag. Red dots correspond to states with significant contribution from silicene. Reproduced with permission from Ref.[39]. Copyright 2013 American Physical Society.

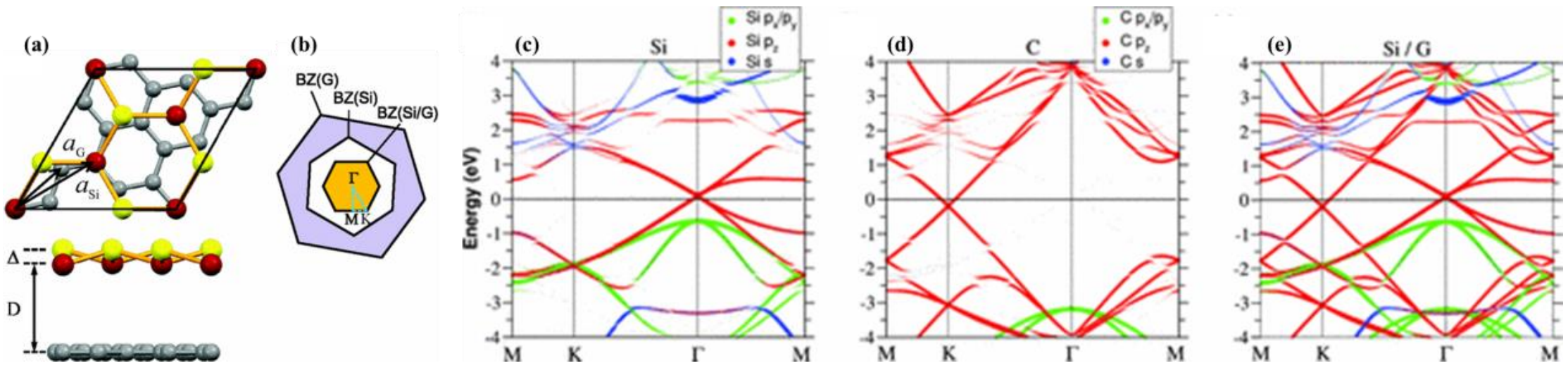

**Figure 6.** (a) Top and side views of the atomic structure of $\sqrt{3}\times\sqrt{3}$ silicene and $\sqrt{7}\times\sqrt{7}$ graphene, where red and yellow atoms represent the two sublattices of Si atoms separated by vertically 0.62 Å, and the grey spheres represent carbon atoms in the graphene layer at a separation of 3.3 Å from the lower silicon layer. The lattice vectors $a_{\mathrm{Si}}$ and $a_{\mathrm{G}}$ of the (1×1) unit cell of silicene and graphene, respectively, have a relative rotational angle of 10.9° between them. (b) The first Brillouin zones of the combined system (orange), 1×1 silicene (white), and 1×1 graphene (purple) plotted. Band structure of $\sqrt{3}\times\sqrt{3}$ silicene and $\sqrt{7}\times\sqrt{7}$ graphene : (c) the projected states on Si are highlighted; (d) the projected states on C are highlighted; and (e) the projected bands in (c) and (d) are combined. The substrate-induced gap is about 26 meV for Si (Γ) and 2 meV for graphene (*K*), respectively. Reproduced with permission from Ref.[27]. Copyright 2013 American Physical Society.

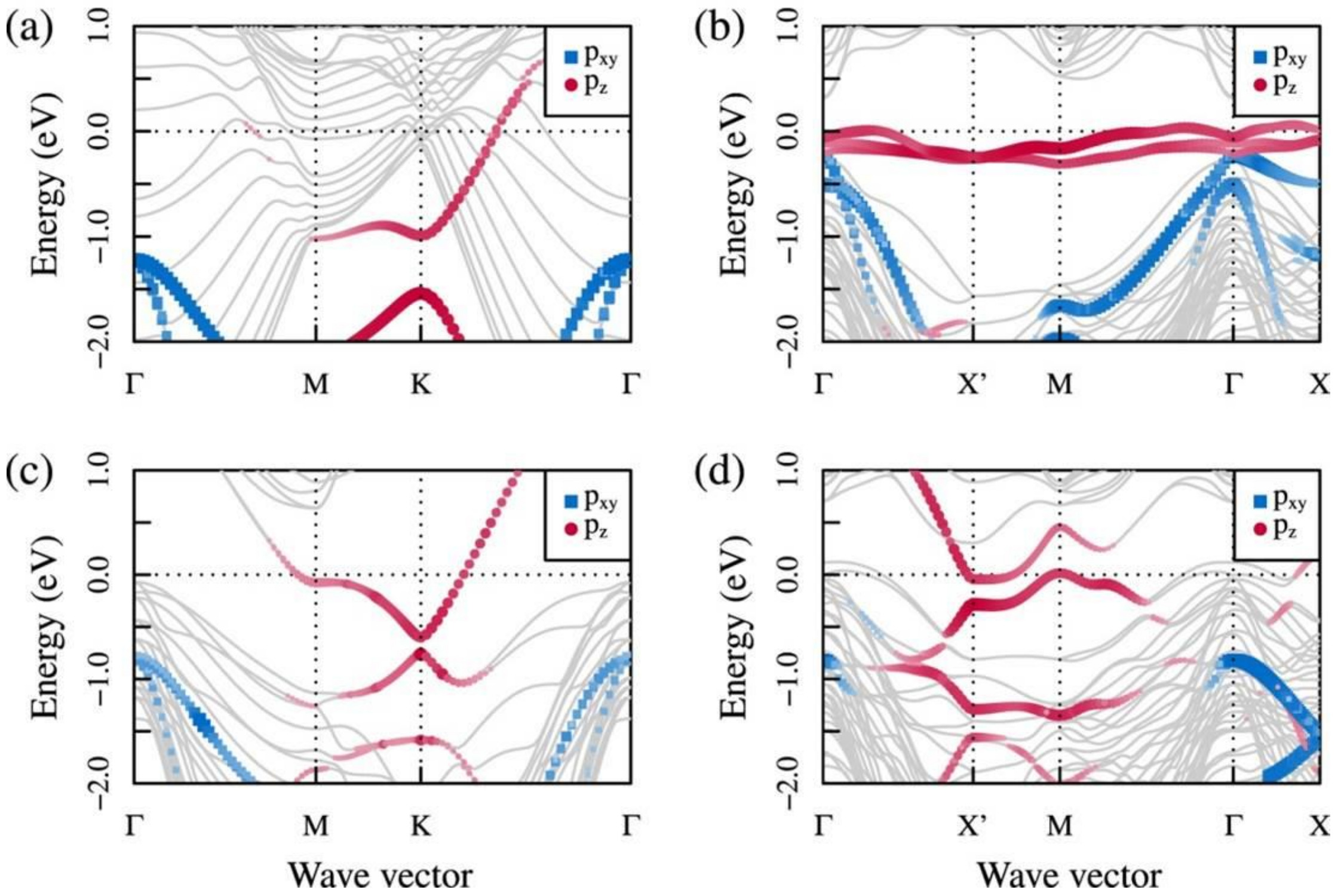

**Figure 7.** Band structures of silicene on (a) Ca(111)1 × 1, (b) Ca(0.5ML)/Si(111)2 × 1, (c) Ca(1.0ML)/Si(111)1 × 1 and (d) Ca(1.5ML)/Si(111)2 × 1 surfaces. The silicene $p_z$ (silicene $p_{xy}$) character of the bands is indicated by red dots (blue squares) of varying size. All bands without silicene character are given by grey lines. Reproduced with permission from Ref.[22]. Copyright 2014 Institute of Physics.

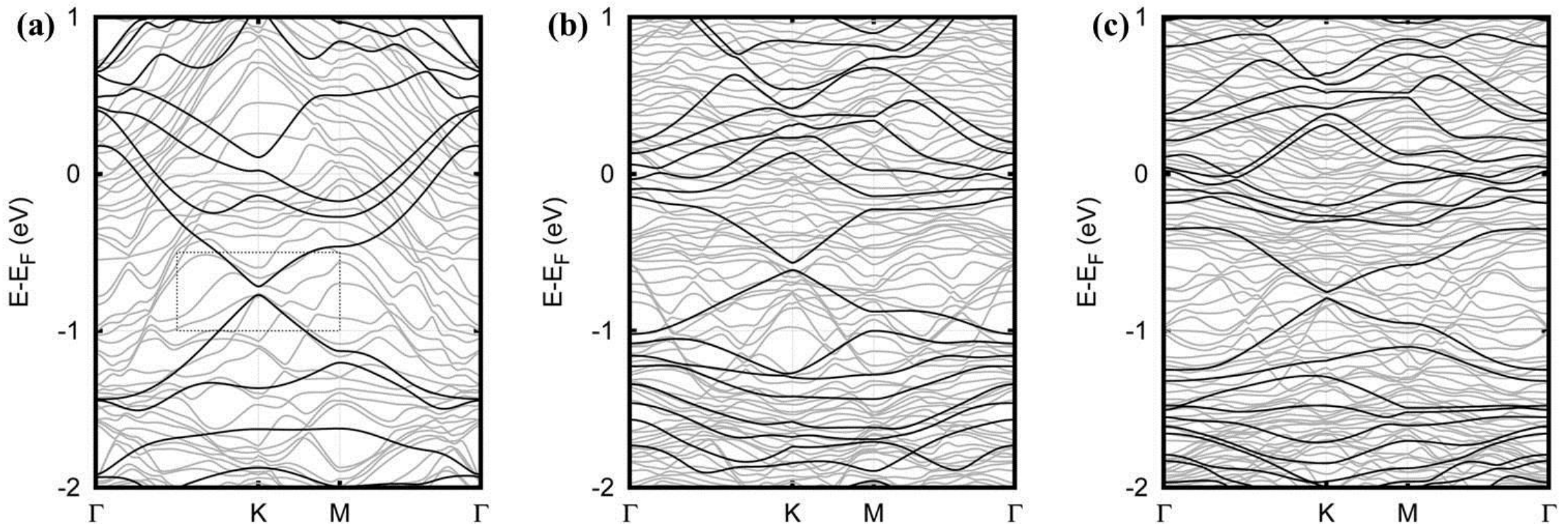

**Figure 8.** Band structures of the silicene superstructures (a) $\sqrt{7}\times\sqrt{7}$ R19.1°, (b) $\sqrt{13}\times\sqrt{13}$ R13.9° and (c) 4×4 on a Pb(111) surface (light) and the standalone silicene layer with atoms fixed in their relaxed positions on the surface (black). The rectangle in panel (a) marks region of the band structure subject to further analysis. Reproduced with permission from Ref.[23]. Copyright 2015 Royal Society of Chemistry.

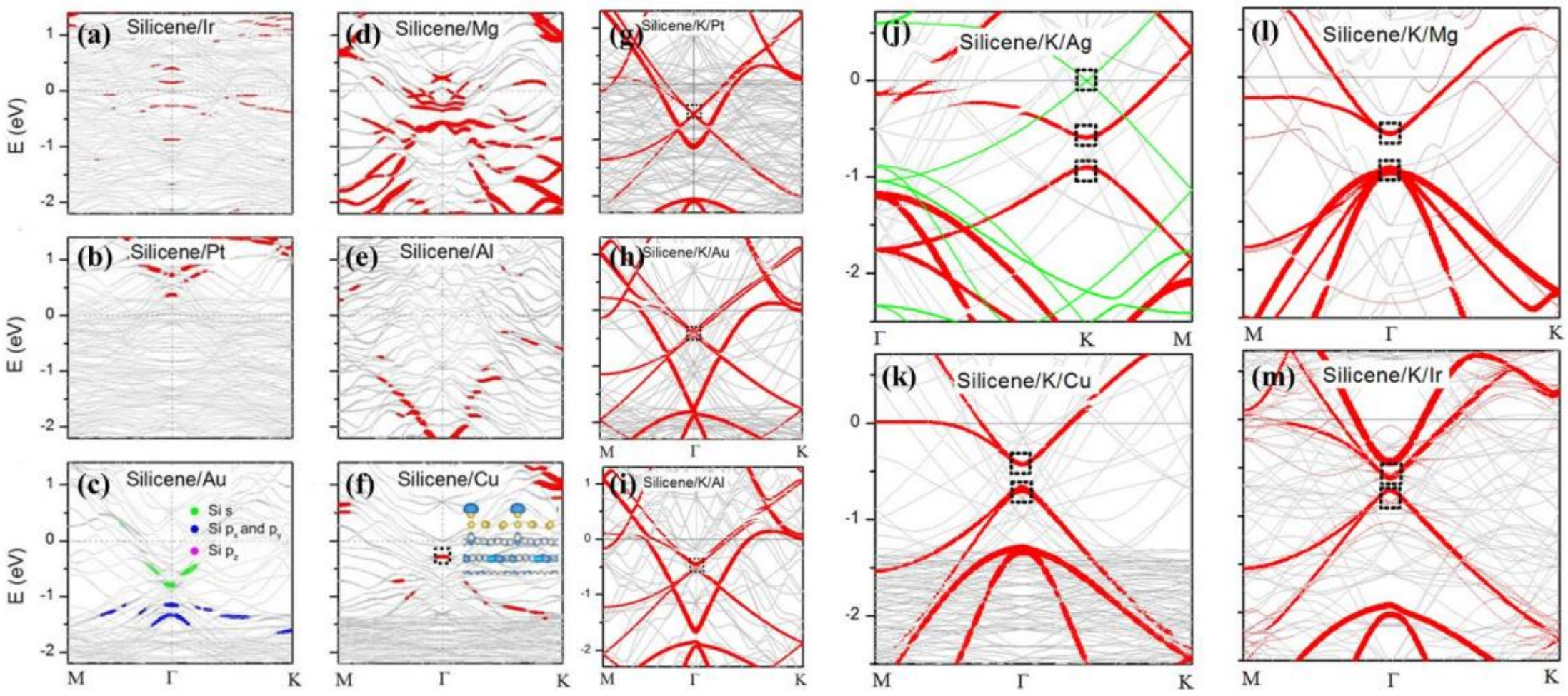

**Figure 9.** (a)-(f) Band structures of epitaxial $\sqrt{3}\times\sqrt{3}$ silicene on various metallic Ir, Pt, Au, Mg, Al, Cu surfaces. (g)-(m) Band structures of epitaxial silicene on metals with K atoms intercalation. $E_f$ is set to zero. The red color indicates the states contributed by the Si atoms, and its thickness is proportional to the Si atom character. Black squares in (g)-(m) indicate those states contributed by the π (π*) bands of silicene. Reproduced with permission from Ref.[46]. Copyright 2014 Nature Publication Group.

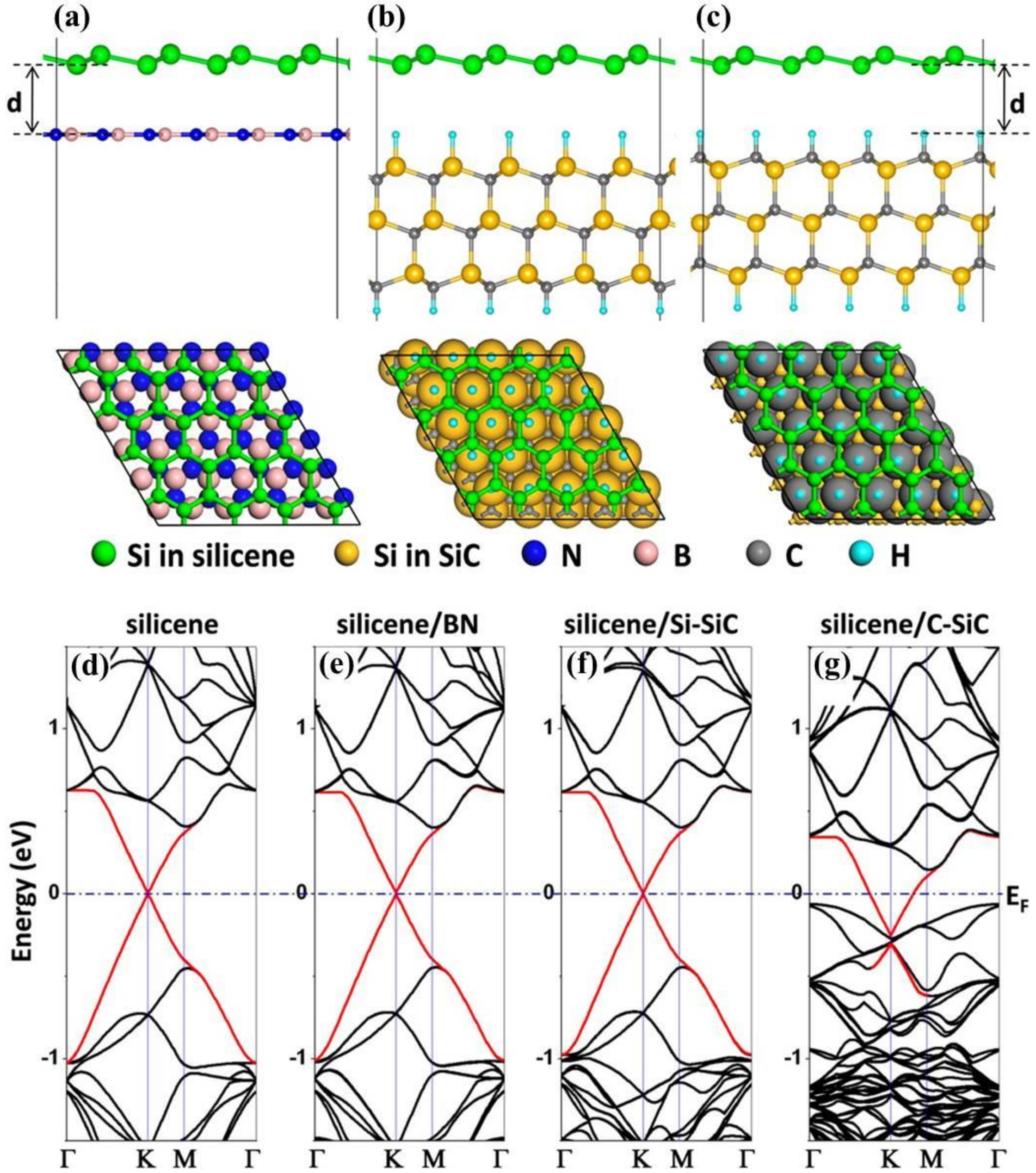

**Figure 10.** Side (upper) and top (lower) views of the atomic configurations of (a) silicene/BN, (b) silicene/Si−SiC, (c) silicene/ C−SiC hybrid systems. The corresponding electronic band structures of (d) silicene, (e) silicene/BN, (f) silicene/Si−SiC, (g) silicene/C−SiC. The red lines highlight the π-bands of silicene in different configurations. Reproduced with permission from Ref.[24]. Copyright 2013 American Chemical Society.

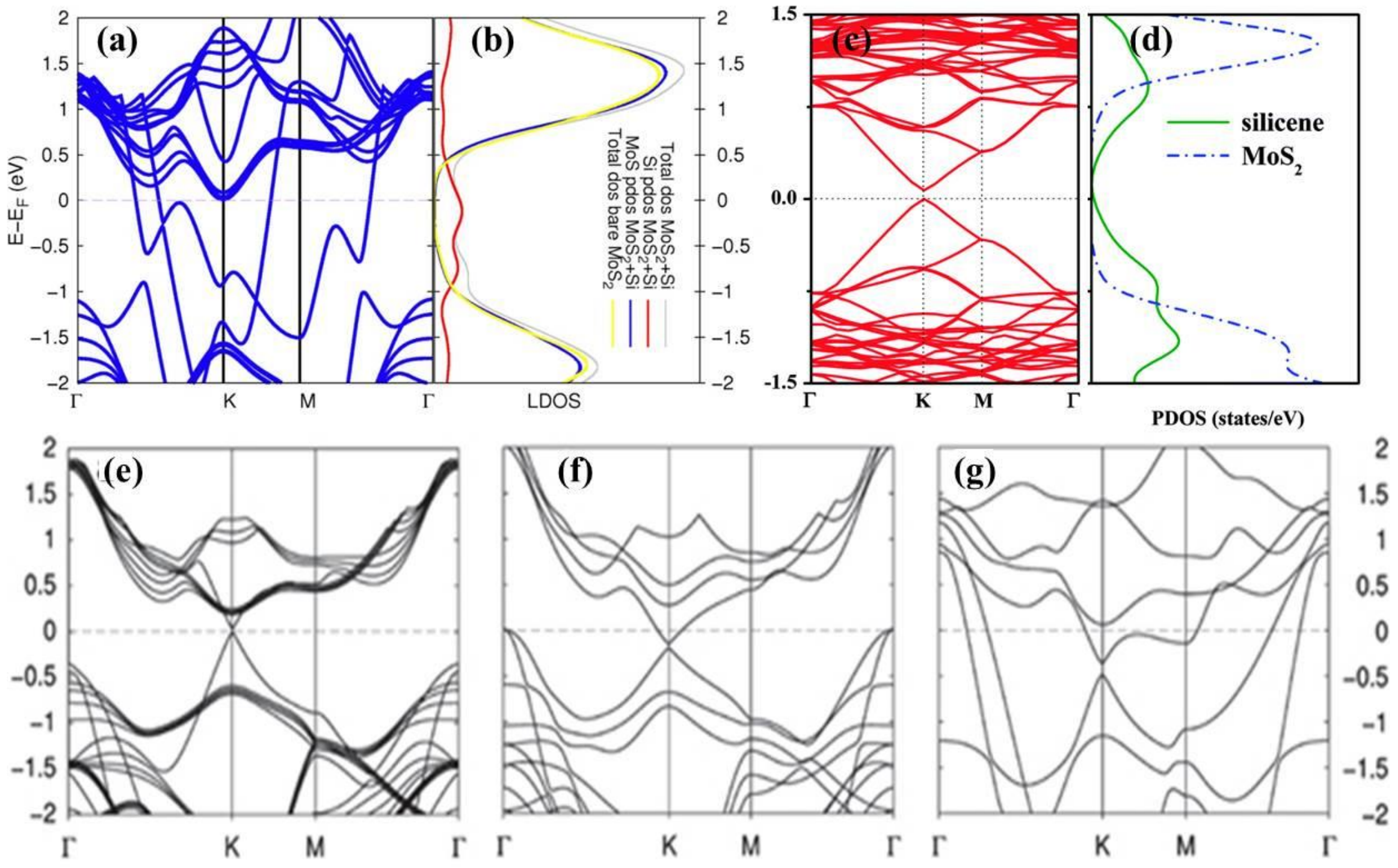

**Figure 11.** Electronic band structure (a) and local density of states (b) of silicene on bulk $MoS_2$. Electronic band structure (c) and partial density of states (d) of silicene on monolayer $MoS_2$. Band structures of silicene on (e) bulk $MoTe_2$, (f) monolayer $MoTe_2$, and (g) $MoSe_2$. Reproduced with permissions from Ref.[25, 59]. Copyright 2014 Institute of Physics and Royal Society of Chemistry.

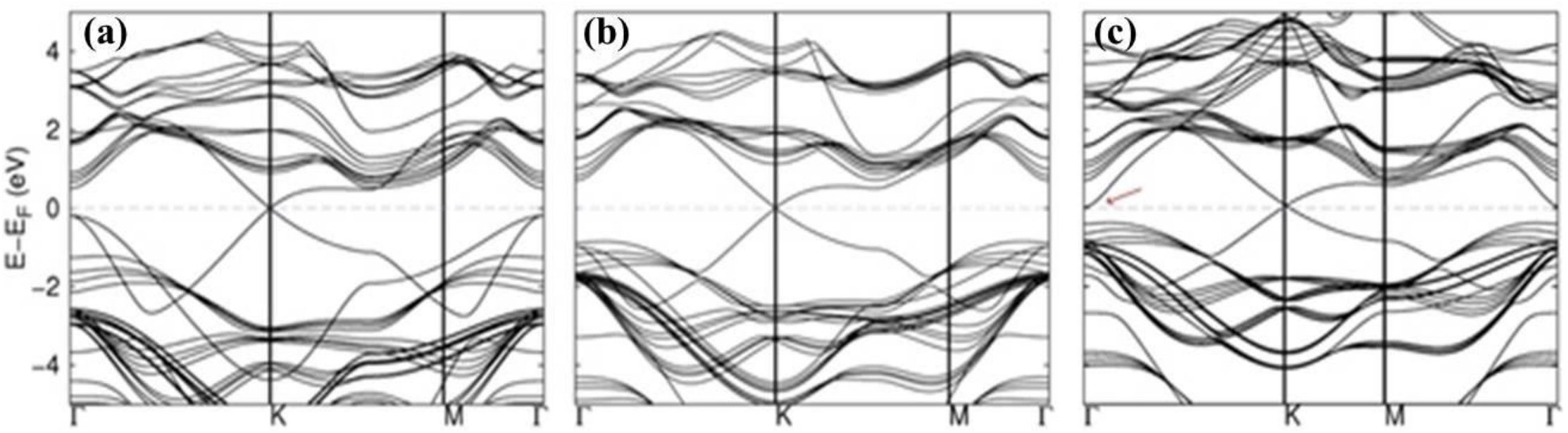

**Figure 12.** Electronic band structures of the silicene layer on (a) bulk GaS, (b) GaSe, and (c) GaTe. The red arrow in (c) indicates the band crossing $E_f$. Reproduced with permission from Ref.[25]. Copyright 2014 Institute of Physics.

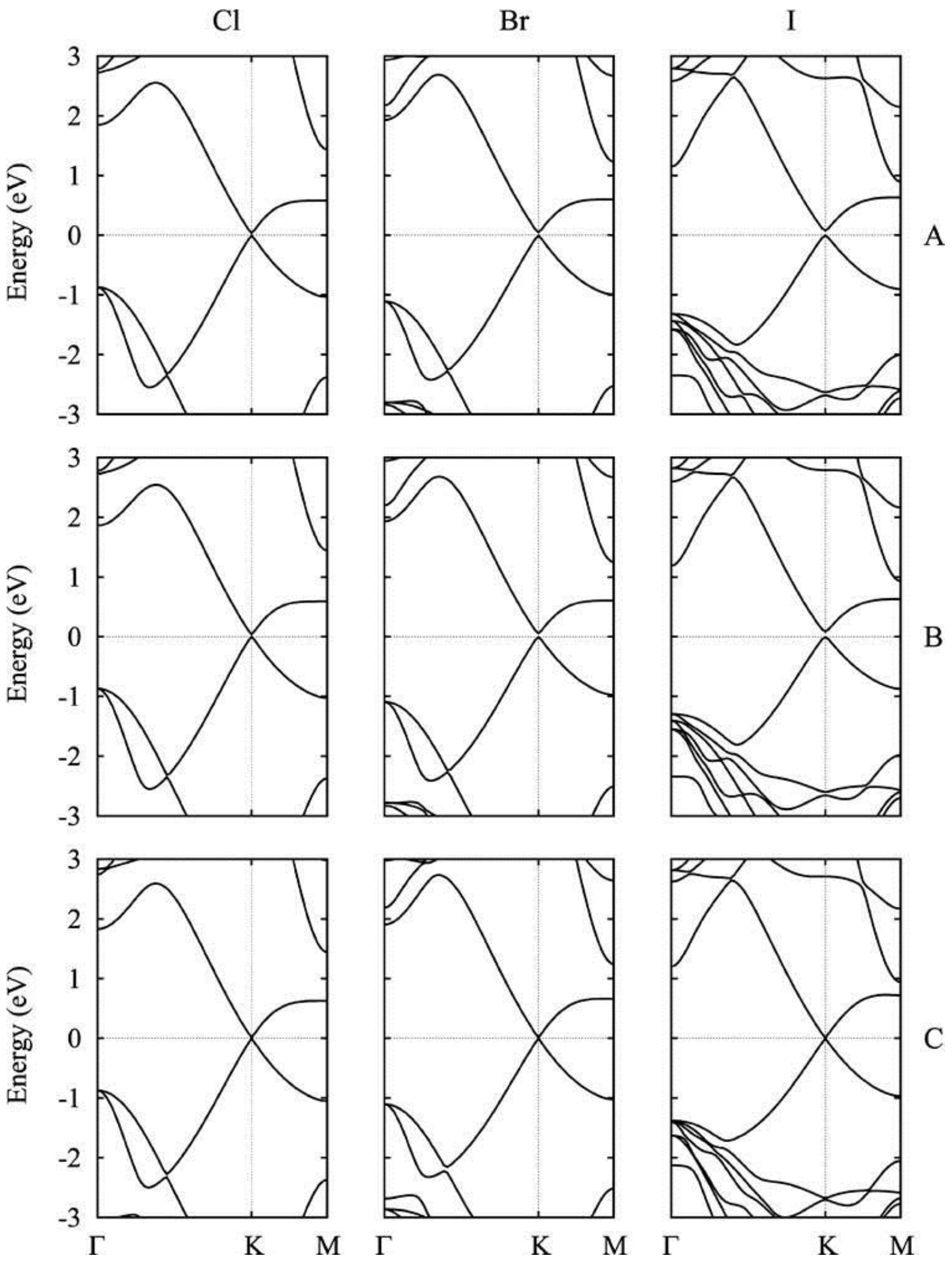

**Figure 13.** Band structures of silicene on $MgX_2$(0001) (X = Cl, Br, and I) for configurations A−C. The $E_f$ is set to zero. Reproduced with permission from Ref.[61]. Copyright 2014 American Chemical Society.